\documentclass[aps,pre,showpacs,twocolumn,floats]{revtex4}
\usepackage{amssymb}

\usepackage{amsmath}
\usepackage{graphicx,psfig}
\usepackage{dcolumn}
\usepackage{bm}

\input{tcilatex}

\begin{document}

\title{Relaxation of the Electron Spin in Quantum Dots Via One- and Two-Phonon
Processes}
\author{C. Calero, E. M. Chudnovsky, and D. A. Garanin}
\affiliation{Department of Physics and Astronomy, Lehman College,
City University of New York \\ \mbox{250 Bedford Park Boulevard
West, Bronx, New York 10468-1589, U.S.A.}}
\date{\today}

\begin{abstract}
We have studied direct and Raman processes of the decay of
electron spin states in a quantum dot via radiation of phonons
corresponding to elastic twists. Universal dependence of the spin
relaxation rate on the strength and direction of the magnetic
field has been obtained in terms of the electron gyromagnetic
tensor and macroscopic elastic constants of the solid.
\end{abstract}

\maketitle
\section{Introduction}

Electron spin relaxation in solids is related to such important
applied problems as electron spin resonance and quantum computing.
Many interactions contribute to the relaxation time of the
electron spin in a semiconductor quantum dot. In principle, all of
them can be eliminated (such as, e.g., impurities, nuclear spins,
etc.) but the interaction with phonons cannot. Thus, spin-phonon
interaction provides the most fundamental upper bound on the
lifetime of electron spin states. The existing methods of
computing electron spin-phonon rates in semiconductors rely upon
phenomenological models of spin-orbit interaction, see, e.g.,
Refs.
\cite{Dresselhaus,Rashba,Hasegawa,Roth,Khaetskii,Glavin,Rashba1,Efros,Loss,Tahan}.
These models contain unknown constants that must be obtained from
experiment. Meantime, the spin-orbit coupling determines the
difference of the electron g-tensor from the unit tensor. The
question then arises whether the effect of the spin-orbit coupling
on spin-phonon relaxation can be expressed via the difference
between the electron gyromagnetic tensor $g_{\alpha \beta}$ and
the vacuum tensor $g_0\delta_{\alpha \beta}$. Since $g_{\alpha
\beta}$ can be measured independently, this would enable one to
compare the computed relaxation rates with experiment without any
fitting parameters. In this paper we show that this can be done
under certain reasonable simplifying assumptions. We obtain
spin-phonon relaxation rates due to direct one-phonon processes
and due to two-phonon Raman processes.

Zeeman interaction of the electron with an external magnetic
field, ${\bf H}$, is given by the Hamiltonian
\begin{equation}\label{Zeeman}
\hat{H}_Z = -{\mu}_B \, g_{\alpha \beta}\, S_{\alpha} H_{\beta}\;,
\end{equation}
where $\mu_B$ is the Bohr magneton and ${\bf S} = {\bf \sigma}/2$
is the dimensionless electron spin, with ${\sigma}_{\alpha}$ being
Pauli matrices. One can choose the axes of the coordinate system
along the principal axes of the tensor $g_{\alpha \beta}$. Then,
$g_{\alpha \beta} = g_{\alpha} \delta_{\alpha \beta}$.
Perturbation of Eq.\,(\ref{Zeeman}) by phonons has been studied in
the past \cite{Hasegawa,Roth,Glavin} by writing all terms of the
expansion of $g_{\alpha \beta}$ on the strain tensor, $u_{\alpha
\beta }$, permitted by symmetry. This gives spin-phonon
interaction of the form $A_{\alpha \beta \gamma \rho }u_{\alpha
\beta }{S}_{\gamma}B_{\rho}$ with unknown coefficients $A_{\alpha
\beta \gamma \rho}$. To avoid this uncertainty we limit our
consideration to local rotations generated by transverse phonons.
The argument for doing this is three-fold. Firstly, the rate of
the transition accompanied by the emission or absorption of a
phonon is inversely proportional to the fifth power of the sound
velocity \cite{Abragam}. Since the velocity of the transverse
sound is always smaller than the velocity of the longitudinal
sound \cite{LL-elasticity}, the transverse phonons must dominate
the transitions. Secondly, we notice that interaction of the
electron spin with a local elastic twist generated by a transverse
phonon does not contain any unknown constants. Consequently, it
gives parameter-free lower bound on the spin relaxation. Thirdly,
for a dot that is sufficiently rigid to permit only tiny local
rotations as a whole under an arbitrary elastic deformation, the
emission or absorbtion of a quantum of the elastic twist is the
only spin-phonon relaxation channel.

The angle of the local rotation of the crystal lattice in the
presence of the deformation, ${\bf u}({\bf r})$, is given by
\cite{LL-elasticity}
\begin{equation}\label{phi}
\delta{\bf \phi} = \frac{1}{2}{\bf \nabla}\times {\bf u}\;.
\end{equation}
As stated above, the gyromagnetic tensor $g_{\alpha \beta}$ is
determined by the local environment of the quantum dot. In the
presence of long wave deformations of the lattice, the whole
environment is rotated so that the gyromagnetic tensor becomes
\begin{equation}\label{gyro-rot}
g_{\alpha \beta} =  {\mathbb{R}}_{\alpha \alpha'}
{\mathbb{R}}_{\beta \beta'} g_{\alpha' \beta'}\,,
\end{equation}
where ${\mathbb{R}}_{\alpha \beta}$ is the $3\times 3$ rotation
matrix given by $\delta {\bf \phi}({\bf r})$. One can thus write
the total Hamiltonian in the form
\begin{equation}\label{totalHamiltonian}
\hat{\mathcal{H}} = \hat{\mathcal{H}}_{Z}^{\prime} +
\hat{\mathcal{H}}_{ph}\,,
\end{equation}
where
\begin{equation}\label{HZph}
\hat{\mathcal{H}}_{Z}^{\prime} = -\mu_B g_{\alpha' \beta'} \left(
{\mathbb{R}}_{\alpha' \alpha}^{-1} S_{\alpha} \right)\left(
{\mathbb{R}}_{\beta' \beta}^{-1} H_{\beta} \right)\,.
\end{equation}
and $\hat{\mathcal{H}}_{ph}$ describes free harmonic phonons. In
the above formulae, $\delta {\bf \phi}$ must be understood as an
operator. Indeed, from the canonical quantization of phonons and
Eq. (\ref{phi}) one obtains
\begin{equation}\label{deltaphiquant}
\delta \bf {\phi }  =  \sqrt{\frac{\hbar }{8 \rho V}}\sum_{\mathbf{k}%
\lambda }\frac{\left[ i\mathbf{k}\times \mathbf{e}_{\mathbf{k}\lambda }%
\right] e^{i\mathbf{k\cdot r}}}{\sqrt{\omega _{\mathbf{k}\lambda
}}}\left( a_{\mathbf{k}\lambda }+a_{-\mathbf{k}\lambda }^{\dagger
}\right)
\end{equation}
where $\rho$ is the mass density, $V$ is the volume of  the
crystal, $\mathbf{e}_{\mathbf{k}\lambda }$ are unit polarization
vectors, $ \lambda =t_{1},t_{2},l$ denotes polarization, and
$\omega _{k\lambda }=v_{\lambda }k$ is the phonon frequency.

\section{Direct processes}
In order to account for spin transitions accompanied by the
emission or absorption of one phonon one needs to consider terms
up to first order in phonon amplitudes. Therefore, with the help
of the expansion of the rotation matrix to the first order in
$\delta {\bf \phi}$, ${\mathbb{R}}_{\alpha \beta}=\delta _{\alpha
\beta}-\epsilon _{\alpha \beta \gamma}\delta \phi _{\gamma}$, we
obtain the full Hamiltonian
\begin{equation}\label{fullHamiltonian}
\hat{\mathcal{H}} = \hat{\mathcal{H}}_0 +
\hat{\mathcal{H}}_{s-ph}\,, \qquad \hat{\mathcal{H}}_0 =
\hat{\mathcal{H}}_Z + \hat{\mathcal{H}}_{ph}\,,
\end{equation}
where
\begin{equation}\label{Hsph}
\hat{\mathcal{H}}_{s-ph} = -\mu_B\epsilon_{\alpha \beta
\gamma}(g_{\alpha} - g_{\beta}) H_{\beta}
\delta\phi_{\gamma}S_{\alpha}\,.
\end{equation}
Spin-phonon transitions occur between the eigenstates of
$\hat{\mathcal{H}}_0$, which are direct products of spin and
phonon states
\begin{equation}\label{eigenstates1}
\left|\Psi_{\pm}\right\rangle = \left|\psi_{\pm}\right\rangle
\otimes \left|\phi_{\pm}\right\rangle\,.
\end{equation}
Here $ \left|\psi_{\pm}\right\rangle   $ are the eigenstates of
$\hat{\mathcal{H}}_Z$ with energies $E_{\pm}$ and
$\left|\phi_{\pm}\right\rangle\ $ are the eigenstates of
$\hat{\mathcal{H}}_{\mathrm{ph}}$ with energies
$E_{\mathrm{ph}\pm}$. Spin-phonon processes conserve energy,
$E_{+} + E_{\mathrm{ph} +} = E_{-} + E_{\mathrm{ph} -}$. For
$\hat{\mathcal{H}}_{s-ph}$ of Eq.\,(\ref{fullHamiltonian}), the
states $\left|\phi_{\pm}\right\rangle $ differ by one emitted or
absorbed phonon with wave vector ${\bf k}$. We will use the
following designations
\begin{equation}\label{phonon states}
\left|\phi_{+}\right\rangle \equiv
\left|n_{\bf{k}}\right\rangle,\; \left|\phi_{-}\right\rangle
\equiv \left|n_{\bf{k}} +1\right\rangle.
\end{equation}
To obtain the relaxation rate one can use the Fermi golden rule.
The matrix element corresponding to the decay of the spin
$|\Psi_+\rangle \rightarrow |\Psi_-\rangle$ can be evaluated with
the help of Eqs.\,(\ref{Hsph}) and (\ref{deltaphiquant}) we obtain
\begin{eqnarray}
&&\langle \Psi_-| \hat{\mathcal{H}}_{s-ph}|\Psi_+\rangle  =
\frac{\hbar}{\sqrt{V}} \sum_{{\bf k} \lambda} V_{{\bf k}\lambda}
\langle n_{\bf k'}+1| a_{{\bf k}\lambda} + a_{-{\bf k}
\lambda}^{\dagger} |n_{\bf k'}\rangle \nonumber \\
&&V_{{\bf k}\lambda} \equiv \frac{e^{i{\bf k}\cdot {\bf
r}}}{\sqrt{8\rho \hbar \omega_{{\bf k}\lambda}}}\,{\bf K}\cdot
[{\bf k}\times {\bf e}_{{\bf k}\lambda}]\label{V}
\end{eqnarray}
where the components of vector ${\bf K}$ are given by
\begin{equation}\label{K}
K_{\gamma} \equiv -\mu_B \epsilon_{\alpha \beta \gamma}
(g_{\alpha} - g_{\beta})H_{\beta}\langle \psi_-|S_{\alpha}
|\psi_+\rangle\,.
\end{equation}
Note that only the transverse phonons are considered in the
summations. For the direct process, the decay rate, $\Gamma_{D}$,
is, then,
\begin{equation}
\Gamma_{D} = \Gamma_0 \coth\left(\frac{\hbar
\omega_0}{2k_BT}\right)\,.
\end{equation}
where
\begin{equation}
\hbar \omega_0 \equiv E_+ - E_- = \mu_B \left(\sum_{\gamma}
g_{\gamma}^2 H_{\gamma}^2 \right)^{1/2}
\end{equation}
is the distance between the two spin levels, and
\begin{equation}
\Gamma_0 = \frac{1}{V} \sum_{{\bf k} \lambda} |V_{{\bf k}
\lambda}|^2 2\pi \delta(\omega_{{\bf k} \lambda} - \omega_0)\,.
\end{equation}
Using Eq.\,(\ref{V}) and replacing $\sum_{\bf k}$ by
$(V/(2\pi)^3)\int d^3k$, one obtains
\begin{equation}
\Gamma_0 = \frac{1}{12\pi \hbar}\frac{|{\bf K}|^2 \omega_0^3}{\rho
v_t^5}\,,
\end{equation}
where $v_t$ is the velocity of the transverse sound. A
straightforward calculation of the spin matrix elements yields
\begin{eqnarray}
|{\bf K}|^2 & = & \frac{\mu_B^2}{8}\sum_{\alpha \beta = x, y,
z}(g_{\alpha} - g_{\beta})^2\nonumber \\
&\times& \left[B_{\alpha}^2 + B_{\beta}^2 - \frac{(g_{\alpha} +
g_{\beta})^2H_{\alpha}^2H_{\beta}^2}{\sum_{\gamma}(g_{\gamma}H_{\gamma})^2}\right].
\end{eqnarray}
Then, the decay rate can be written in the final form
\begin{equation}
\Gamma_{D} = \frac{\hbar}{3\pi \rho}\left(\frac{\mu_B H}{\hbar
v_t}\right)^5 F_T({\bf n})\,,
\end{equation}
where ${\bf n} \equiv {\bf H}/H$ and
\begin{eqnarray}\label{FT}
&& F_T({\bf n}) = \left(\sum_{\gamma} g_{\gamma}^2n_{\gamma}^2
\right)^{3/2}\coth \left[ \frac{\mu _{B}H}{2k_{B}T}
\left( \sum_{\gamma }g_{\gamma }^{2}n_{\gamma }^{2}\right) ^{1/2}\right]\nonumber \\
&&\times\frac{1}{32}\sum_{\alpha \beta}(g_{\alpha} -
g_{\beta})^2\left[n_{\alpha}^2 + n_{\beta}^2 - \frac{(g_{\alpha} +
g_{\beta})^2n_{\alpha}^2n_{\beta}^2}{\sum_{\gamma}
(g_{\gamma}n_{\gamma})^2} \right]\,.\nonumber \\
\end{eqnarray}
If the field is directed along the $Z$-axis, Eq.\,(\ref{FT})
simplifies to
\begin{equation}
F_T({\bf e}_z) = \frac{g_z^3}{32}[(g_z-g_x)^2 + (g_z-g_y)^2]\coth
\left( \frac{g_{z}\mu _{B}B}{2k_{B}T}\right).
\end{equation}
When components of $g_{\alpha\beta}$ are of order unity and $k_BT
\gtrsim \mu_B H$, then $\Gamma_D\sim (k_B T/\hbar)(\mu_BH/E_t)^4$,
where we have introduced $E_t \equiv (\hbar^3 \rho v_t^5)^{1/4}
\sim 10^2$K.

\section{Raman processes}
Spin-lattice relaxation by Raman scattering is a two phonon
mechanism consisting of a spin transition accompanied by the
absorption of a phonon and the emission of another phonon of
different frequency. In spite of being a second order process, its
contribution can be very important, since the phase space of the
phonons triggering the transition is not limited to the distance
between the levels as in the direct case. To describe such
processes, we need to consider terms up to second order in phonon
amplitudes in the Hamiltonian. To this end we expand the
$3\times3$ rotation matrix to second order in $\delta {\bf \phi}$
\begin{equation}
{\mathbb{R}}_{\alpha \beta} = \delta_{\alpha \beta} -
\epsilon_{\alpha \beta \gamma} \delta \phi_{\gamma} +
\frac{1}{2}\left[ \delta \phi_{\alpha}\delta \phi_{\beta} -
\delta_{\alpha \beta}(\delta {\bf \phi})^2\right]\,
\end{equation}
and obtain the full Hamiltonian from
Eqs.\,(\ref{totalHamiltonian}) and (\ref{HZph})
\begin{equation}\label{fullHamiltonian2}
\hat{\mathcal{H}} = \hat{\mathcal{H}}_0 +
\hat{\mathcal{H}}_{s-ph}\,, \qquad \hat{\mathcal{H}}_{s-ph} =
\hat{\mathcal{H}}_{s-ph}^{(1)} + \hat{\mathcal{H}}_{s-ph}^{(2)}\,,
\end{equation}
where $\hat{\mathcal{H}}_{s-ph}^{(1)}$ is given by Eq.
(\ref{Hsph}) and
\begin{eqnarray}\label{Hsph2}
\hat{\mathcal{H}}_{s-ph}^{(2)} &=& -\frac{1}{2}\mu_B (g_{\alpha} +
g_{\beta})S_{\alpha}H_{\beta}\delta \phi_{\alpha}\delta
\phi_{\beta} \nonumber \\ & - & \mu_B g_{\alpha}({\bf {S}}\times
\delta{\bf \phi})_{\alpha}({\bf H} \times \delta{\bf
\phi})_{\alpha}\,.
\end{eqnarray}
Again, we will study the spin-phonon transitions between the
eigenstates of $\hat{\mathcal{H}}_0$,
$\left|\Psi_{\pm}\right\rangle = \left|\psi_{\pm}\right\rangle
\otimes \left| \phi_{\pm} \right \rangle$. Here,
again,$|\psi_{\pm}\rangle$ are the eigenstates of
$\hat{\mathcal{H}}_Z$ with energies $E_{\pm}$ and
$|\phi_{\pm}\rangle$ the eigenstates of $\hat{\mathcal{H}}_{ph}$
with energies $E_{ph\pm}$. We consider Raman processes, in which a
phonon with wave vector ${\bf k}$ is absorbed and a phonon with a
wave vector ${\bf q}$ is emitted. We will use the following
designations for the phonon states
\begin{equation}
|\phi_+\rangle \equiv |n_{\bf k}, n_{\bf q}\rangle , \qquad
|\phi_-\rangle \equiv |n_{\bf k}-1, n_{\bf q}+1\rangle\,.
\end{equation}
To obtain the relaxation rate of the transition $|\Psi_+\rangle
\rightarrow |\Psi_-\rangle$ one needs to evaluate the matrix
element of the process, which is the sum of the matrix element
with $\hat{\mathcal{H}}_{s-ph}^{(2)}$ and that with
$\hat{\mathcal{H}}_{s-ph}^{(1)}$ in the second order: $M_R =
M_R^{(2)} + M_R^{(1+1)}$, where
\begin{equation}
M_R^{(2)} = \left \langle \Psi_- \right|
\hat{\mathcal{H}}_{s-ph}^{(2)} \left| \Psi_+\right \rangle
\end{equation}
and
\begin{eqnarray}
&& M_{R}^{(1+1)}=\sum_{\xi = \pm }\frac{\left\langle \Psi _{-}\left| \hat{\mathcal{H}}_{%
\mathrm{s-ph}}^{(1)}\right| \Psi _{\xi }\right\rangle \left\langle
\Psi
_{\xi }\left| \hat{\mathcal{H}}_{\mathrm{s-ph}}^{(1)}\right| \Psi _{+}\right\rangle }{%
E_{+}+\hbar \omega _{\mathbf{k}}-E_{\xi }}  \nonumber \\
&&+\sum_{\xi=\pm }\frac{\left\langle \Psi _{-}\left| \hat{\mathcal{H}}_{\mathrm{s-ph}%
}^{(1)}\right| \Psi _{\xi }\right\rangle \left\langle \Psi _{\xi
}\left| \hat{\mathcal{H}}_{\mathrm{s-ph}}^{(1)}\right| \Psi
_{+}\right\rangle }{E_{+}-E_{\xi }-\hbar \omega _{\mathbf{q}}}.
\label{MRaman11}
\end{eqnarray}
The intermediate phonon states are $|n_{\bf k} -1, n_{\bf q}
\rangle$ in the first term and $|n_{\bf k}, n_{\bf q} +1\rangle$
in the second term.

Raman processes may dominate over direct processes for a small
energy difference between the spin states, $\hbar \omega_0 \ll k_B
T$. Thus we will consider terms to the lowest order in H. For
simplicity we will study the case where the field is directed
along the $Z$-axis. Then the matrix element becomes,
\begin{eqnarray}
M_R & = & M_R^{(2)} = \frac{\mu_B H}{4}\big[(2g_y - g_x -
g_z)\tilde{M}_{ph-R}^{x z}  \nonumber \\
& - & i (2g_x - g_y - g_z)\tilde{M}_{ph-R}^{y z} \big]\,,
\end{eqnarray}
where $\tilde{M}_{ph-R}^{\alpha \beta} = {M}_{ph-R}^{\alpha \beta}
+ {M}_{ph-R}^{\beta \alpha}$ with
\begin{eqnarray}
M_{ph-R}^{\alpha \beta }& = &\left\langle n_{\mathbf{q}}+1\left|
\delta \phi _{\alpha }\right| n_{\mathbf{q}}\right\rangle \left\langle n_{%
\mathbf{k}}-1\left| \delta \phi _{\beta }\right|
n_{\mathbf{k}}\right\rangle \nonumber \\
&=&\frac{\hbar ^{2}}{8\rho V}\frac{\left[ \mathbf{%
k}\times \mathbf{e}_{\mathbf{k}\lambda _{\mathbf{k}}}\right]
_{\alpha }\left[ \mathbf{q}\times \mathbf{e}_{\mathbf{q}\lambda
_{\mathbf{q}}}\right] _{\beta
}}{\sqrt{\hbar \omega _{\mathbf{k}\lambda _{\mathbf{k}}}\hbar \omega _{%
\mathbf{q}\lambda _{\mathbf{q}}}}}\sqrt{\left( n_{\mathbf{q}}+1\right) n_{%
\mathbf{k}}}\nonumber \\
\end{eqnarray}
The Raman rate of the transition $|\Psi_+\rangle \rightarrow
|\Psi_-\rangle$ can be obtained by using the Fermi golden rule,
\begin{eqnarray}
\Gamma_R &= & \left[(2g_y-g_x-g_z)^2 + (2g_x-g_y-g_z)^2
\right]\nonumber \\
& \times & \frac{\pi^3}{3024}\frac{k_B T}{\hbar}
\left(\frac{\mu_BH}{E_t}\right)^2\left(\frac{k_BT}{E_t}\right)^6
\,.
\end{eqnarray}\\
The ratio of Raman and direct rates is of order $\Gamma_R/\Gamma_D
\sim 10^{-2}(k_BT/\mu_BH)^2(k_BT/E_t)^4$. Consequently, Raman
processes dominate over direct processes at high temperature and
low field.\\

This work has been supported by the NSF Grant No. EIA-0310517.

\end{document}